\documentclass[aps,pra,groupedaddress,10pt,twocolumn]{revtex4-1}
\usepackage{amssymb,amsmath}
\usepackage[varg]{txfonts}
\usepackage{graphicx}
\usepackage[colorlinks,linkcolor=blue,citecolor=blue,urlcolor=blue]{hyperref}

\mathchardef\mhyphen="2D

\newcommand{\enquote}[1]{''#1''}

\begin{document}

\title{Optical vortices discern  attosecond time delay in electron emission  from  magnetic sublevels}

\author{J. W\"{a}tzel}
%\email{jonas.waetzel@physik.uni-halle.de}
\author{J. Berakdar}
%\email{jamal.berakdar@physik.uni-halle.de}
\affiliation{Institute for Physics,  Martin-Luther-University Halle-Wittenberg, 06099 Halle, Germany}

\date{\today}
\begin{abstract}
Photoionization from energetically distinct electronic states may have  a relative time delay of tens of attoseconds.
Here we demonstrate that  pulses of optical vortices   allow measuring  such attoseconds delays  from  magnetic sublevels, even from a spherically symmetric target. The difference in the time delay is substantial and exhibits a strong angular dependence. Furthermore, we find an atomic scale variation in the time delays depending on the target orbital position in the laser spot. The findings offer thus a qualitatively new way  for a spatio-temporal sensing of  the magnetic states from which the photoelectrons originate,  with a spatial resolution  way below the diffraction limit of the vortex  beam. Our conclusions follow from analytical considerations based on symmetry, complemented and confirmed with full numerical simulations of the quantum dynamics.
\end{abstract}

\maketitle

\section{Introduction}
The development of attosecond (\emph{as}) optical sources  is a major achievement. Beside  technological applications, attosecond spectroscopy and metrology shed light on new and old fundamental problems that were  hitherto experimentally inaccessible  (cf.  \cite{atto}). An illuminating example is the recent revival of the issues of  time and time delay in tunneling and photoionization  \cite{Pazourek,Torlina,Maquet,Dahlstrom,Landsman,su} which dates  back to  early applications of  quantum mechanics  to study   the  variation of the scattering-amplitude  phase with the wave vector during collision processes \cite{Eisenbud1,Wigner1,joachain1975quantum}. Combining   \emph{as} XUV pulses and  infrared (IR)  streaking fields, the relative time delay of  photo emitted electrons  from different atomic levels  was reported \cite{Schultze}
(for an overview, see e.g. \cite{klunder2011probing,guenot2012photoemission,Pazourek}). The finding triggered  theoretical activities with a varying level of success in reproducing the experiments \cite{kheifets2010delay,nagele2011time,zhang2011streaking,moore2011time,nagele2012time,dahlstrom2012diagrammatic,dahlstrom2013theory,kheifets2013time,dixit2013time,
feist2014time,dahlstrom2014study,watzel2015angular} but  pointing out the role of electronic correlations, the directional dependence of the emitted electrons,  laser fields effects, as well as the influence of resonances and Cooper minima.\\
Here we  draw attention to another  aspect of time-resolved photoelectron chronoscopy when utilizing spatially phase-structured (singular) laser fields,
  i.e. optical vortices \cite{Allen1992,ref3,ref4,ref5,ref6,ref7,Allen,Boyd}. Such beams can transfer orbital angular momentum (OAM) when interacting with matter \cite{FrieseNat1998,ONeil2002,Simpson97,Gahagan96,babiker1994light,andrews2012angular,TL_JW} and have found numerous applications in a number of fields in science \cite{Terriza2007,mair2001entanglement,barreiro2008beating,boyd2011quantum,padgett2011tweezers,furhapter2005spiral,key-16,
torres2011twisted,andrews2011structured,foo2005optical,he1995optical,wang2008creation,hell2007far}.
 The OAM phase front forms a helical shape characterized by $\exp(i m_{_{\rm OAM}}\varphi)$, where $\varphi$ is the azimuthal angle with respect to the propagation direction, and $m_{_{\rm OAM}}$ is an integer winding number  called the topological charge.
 OAM carrying laser beams are  routinely  realized, e.g. as Laguerre-Gaussian (LG) modes.
 Each photon may transfer a quantized OAM of $m_{_{\rm OAM}}\hbar$. Beams with more than $m_{_{\rm OAM}}\hbar=300$ were demonstrated offering the opportunity to access excitations way beyond the limit set by the conventional optical propensity rules  \cite{picon2010photoionization,koksal2012charge}  (degeneracy of  involved states is also important \cite{Watzel2016Driving}). Hence, vortices offer  a  new optical key to access  magnetic sublevels  \cite{Watzel2016Driving} which is the starting idea of this work.\\
We consider an XUV, OAM carrying LG beam ionizing an initially completely symmetric  target such as  Ar atoms or C$_{60}$ molecules. For the experimental feasibility and trapping properties of
   LG beams we refer to \cite{Geneaux} and others \cite{corkumoam,zuerch,Vieira,Zhang,Ashkin,Meschede}. As the amount of transferred OAM  depends strongly on the orbital location in the laser spot   \cite{koksal2012charge,matula2013atomic,afanasev2013off,Watzel2016Driving,picon2010transferring}, and  the predicted time delay is explicitly related to the transferred OAM we predict a spatial resolution on orbitals having time-delay.
\subsection{Background}
The quantity of interest is the dependence of the time delay $\tau$ in photoionization   on the light topological charge.
Usually, in the experiment $\tau$ receives  two contributions $\tau_{\rm W}$ and $\tau_{\rm CLC}$. The Wigner time delay  $\tau_{\rm W}$ is related to the  photoionization process triggered by the XUV pulse. The Coulomb-laser coupling term $\tau_{\rm CLC}$ is akin to photoelectron  motion in the combined Coulomb-streaking field, and hence  is  a setup-dependent quantity \cite{pazourek2013time,dahlstrom2013theory}.  We concentrate here on  Wigner time delay as a system-sensitive quantity \cite{Note1}.
The time delay of a specific subshell is the average contributions over all magnetic sub-states, labelled $m_{ i}$.  For  linearly polarized light the photoionization probabilities of states with $\pm m_{ i}$ are  equal and so are their contributions to the time delay. Also the
 angular dependence of photoelectrons emitted from these states are identical.  Irradiation with   OAM beams allows for transitions involving a change in magnetic quantum numbers by a maximal amount set by $m_{_{\rm OAM}}$. We will show explicitly for Ar atom and a fullerene cluster that a short XUV-OAM pulse (\cite{hernandez2013attosecond,hernandez2014coherent,zhang2015generation})  ionizes preferentially specific magnetic sublevels.
Consequently, in certain directions the photoionization  (and hence the time delay) is  dominated by a specific magnetic sublevel, depending  on the atom position in the beam. The time delay may serve so as a tool to identify the origin of the photoelectron in energy, magnetic state, and space.\\
\section{Model}
In a gauge where the scalar potential vanishes the interaction hamiltonian reads (atomic units are used)
$\hat{H}_{\rm Int}=\boldsymbol{\hat p}\cdot\boldsymbol{A}(\boldsymbol{r},t) +
\boldsymbol{A}(\boldsymbol{r},t)\cdot\boldsymbol{\hat p},$
where $\boldsymbol{\hat p}$ is  the momentum operator and $\boldsymbol{A}$ is the vector potential with polarization vector
$\hat{\epsilon}=(1,i)$, waist $w_0$,
%(we use $w_0=50$\,nm (940\,a.u.),
temporal envelope  $g(t) = \cos[\pi t/nT]^2$, where $T=2\pi/\omega$ is the cycle duration and $n$ is the number of optical cycles.
The explicit form  of $\boldsymbol{A}$   is given in  appendix A.
We focus on Ar  target atom. Technicalities for $C_{60}$ are  in  appendix F.
For Ar the involved  initial states  are captured by  the effective single-particle potential  \cite{muller1999numerical}
%\begin{equation}
$V(r)=-(1+5.4e^{-r}+11.6e^{-3.682r})/r,$
%\end{equation}
which proved useful for similar problems  \cite{toma2002calculation}.  $V(r)$   misses  correlation
effects \cite{higuet2011high}, yet the  energetic position of the Cooper minimum is  reasonably well reproduced  \cite{dahlstrom2014study}.
Note, we consider  photoemitted electrons and  do not study the hole dynamics upon electron removal \cite{barth2014hole,barth2013spin}.
From symmetry considerations when applying OAM beams  a hole current is expected to emerge  which
   orbitally magnetizes  the residual ion.
  Obviously $|\mathbf A|^2$ has a donut shape (appendix A). An Ar atom in the donut center experiences only a weak intensity justifying so
  a non-relativistic perturbative treatment  even for moderately  intense
  fields (the frequency is in the XUV). As the atoms might be distributed over the beam, some would experience the peak intensity.
   However, as shown below
  these atoms show no reaction to a topological charge change and hence no time delay (because the transferrable angular momentum refers to the
  optical axis not the atom center). Hence the two types of atoms, located in the donut center or on its ring, should be distinguishable by measuring the time delay.
    Here, the electric field starts with a zero amplitude at the vortex center reaching, at a distance of 10\,a.u., a peak  amplitude of  1\,a.u. \footnote{Not2}.
   It is important to remark that the time delay dependence on $m_{\rm OAM}$ diminishes rapidly as the atom is displaced (say by 2 a.u.) from the vortex center so that the high intensity region is irrelevant for the  time delay discussed here. We note (and explicitly demonstrate below) that due to the laser donut-type intensity profile,  orbitals with larger extension
    such as for fullerenes  show similar effect as for atoms, but at  intensity  orders of magnitudes smaller than needed for atoms (this difference between Ar and $C_{60}$ is two orders of magnitude). As the effect is of a general nature, we expect the proposed scheme to be useful also for extended systems.\\
Concerning  the emitted electron, its wave function is expressible in a standard way \cite{fedorov1997atomic} as
%\begin{equation}
$\Phi(\boldsymbol{r},t)=\int{\rm d}\boldsymbol{k}\,a(\boldsymbol{k},t) \varphi^{(-)}_{\boldsymbol{k}}(\boldsymbol{r})e^{-i \varepsilon_{\boldsymbol{k}}t}.$
%\end{equation}
 The projection
coefficients $a(\boldsymbol{k},t)$ determine the photoionization amplitude as follows: The  photoinduced emission of an electron  initially  in the bound state labeled  $|\Psi_{ i}(\boldsymbol{r})\rangle$  with energy  $\varepsilon_{ i}$ to the continuum state $\varphi^{(-)}_{\boldsymbol{k}}(\boldsymbol{r})$ with the wave vector $\boldsymbol{k} $  and energy $\varepsilon_{\boldsymbol{k}}=k^2/2$ reads  \cite{fedorov1997atomic}
\begin{align}
a_i(\boldsymbol{k})=-i\int_{-\infty}^{\infty}{\rm d}t'\langle \varphi_{\boldsymbol{k}}^{(-)}|\hat{H}_{int}(t)|\Psi_{ i}\rangle e^{i(\varepsilon_{\boldsymbol{k}}-\varepsilon_{ i})t}.
\label{eq:acoeff}
\end{align}
As established  \cite{fedorov1997atomic,dahlstrom2013theory,amusia2013atomic,amusia1997computation} we  expand  in spherical harmonics $Y_{\ell m}(\Omega)$ as
  $\Psi_{ i}(\boldsymbol{r})=R_{n_{ i}\ell_{ i}}(r)Y_{\ell_{ i}m_{ i}}(\Omega_{\boldsymbol{r}})$ and  $\varphi^{-}_{\boldsymbol{k}}(\boldsymbol{r})=\sum_{\ell=0}^{\infty}\sum_{m=-\ell}^{\ell} i^{\ell}R_{k\ell}(r)e^{-i\delta_{\ell}(k)}Y^{*}_{\ell m}(\Omega_{\boldsymbol{k}})Y_{\ell m}(\Omega_{\boldsymbol{r}})$. The radial wave functions $ R_{k\ell}$ are normalized as $\langle R_{k\ell}|R_{k'\ell}\rangle=\delta\left(\varepsilon_{k}-\varepsilon_{k'}\right)$.
The scattering phases are given by $\delta_{\ell}(k)=\sigma_{\ell}(k)+\eta_{\ell}(k)$ where $\sigma_{\ell}(k)={\rm arg}\left[\Gamma(\ell+1-i/k)\right]$ is the Coulomb phase shift \cite{joachain1975quantum}. The quantity $\eta_{\ell}(k)$ is due to  short range phase interactions \cite{dahlstrom2013theory}.\\
For an analytical model  let  us  consider  \emph{as}   OAM  pulse with  $\hat{\epsilon}=(1,i)^T$,   $m_{\rm OAM}=1$, and the  atom is  in the donut center.  We find $\nabla\cdot\boldsymbol{A}(\boldsymbol{r},t)=0$ and
 $e^{-\rho^2/w_0^2}=1$ for   $w_0=50$\,nm.
 \begin{figure*}[t!]
\centering
\includegraphics[width=15.0cm]{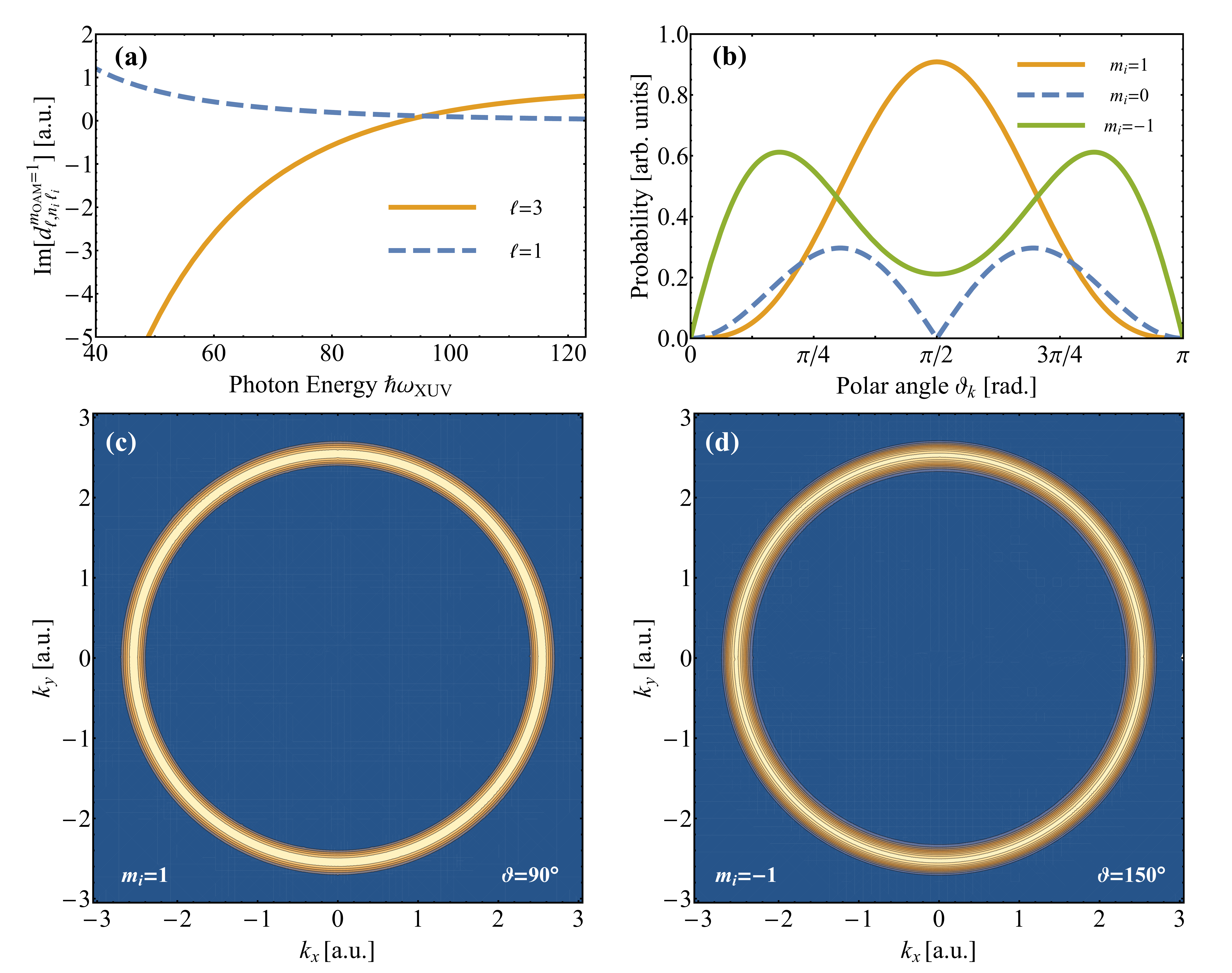}
\caption{(a) Reduced radial matrix elements for  partial wave functions with  orbital angular momenta $\ell=3$ and $\ell=1$. (b) Photoionization probabilities for the three different initial states of  $3p$ subshell in argon. (c) and (d) photoelectron momentum distribution corresponding to $m_{ i}=1$ at $\vartheta_{\boldsymbol{k}}=90^{\circ}$ and $m_{ i}=-1$ at $\vartheta_{\boldsymbol{k}}=150^{\circ}$. The beam has waist  $w_0=50$\,nm  and is $n=10$ optical cycles long.}
\label{fig1}
\end{figure*}
The  angular ($\Omega_{\boldsymbol{k}}=(\vartheta_k,\varphi_k)$) dependent projection coefficients  \eqref{eq:acoeff}  and the reduced  radial matrix elements
$d_{\ell,n_{ i}\ell_{ i}}^{m_{_{\rm OAM}}}$ are given in appendix B. The  photoionization probability
$w_{ i}(\varepsilon_{k},\Omega_{\boldsymbol{k}})=\left|a_{ i}(\boldsymbol{k})\right|^2$  is peaked around the center of energy (COE) given as $\varepsilon_{_{\rm COE}}=\omega+\varepsilon_{ i}$ and $k_{_{\rm COE}}=\sqrt{2\varepsilon_{_{\rm COE}}}$.\\
\section{Results and interpretations}
 The photoionization amplitudes from the magnetic sublevels $m_i$  of the $3p$ subshell have the structure
\begin{equation}
a_i(k_{_{\rm COE}},\Omega_{\boldsymbol{k}})=\begin{cases}\begin{aligned} &S_{m_{_{\rm OAM}}+2,m_{_{\rm OAM}}}Y_{m_{_{\rm OAM}}+2,m_{_{\rm OAM}}} (\Omega_{\boldsymbol{k}}) + \\ &S_{m_{_{\rm OAM}}+2,m_{_{\rm OAM}}}Y_{m_{_{\rm OAM}},m_{_{\rm OAM}}} (\Omega_{\boldsymbol{k}})\end{aligned} &m_i=-1, \\
& \\
S_{m_{_{\rm OAM}}+2,m_{_{\rm OAM}}+1}Y_{m_{_{\rm OAM}}+2,m_{_{\rm OAM}}+1} (\Omega_{\boldsymbol{k}}) &m_i=0, \\
S_{m_{_{\rm OAM}}+2,m_{_{\rm OAM}}+2}Y_{m_{_{\rm OAM}}+2,m_{_{\rm OAM}}+2} (\Omega_{\boldsymbol{k}}) &m_i=1, \end{cases}\nonumber
%\label{eq:scheme}
\end{equation}
%\begin{equation}
\begin{align}S_{\ell,m}=\mathcal{E}_{-}(\varepsilon_{\rm COE})d_{\ell,n_{ i}\ell_{ i}}^{m_{_{\rm OAM}}}i^{-\ell} e^{i\delta_{\ell}(k_{_{\rm COE}})}\begin{pmatrix}\ell&m_{_{\rm OAM}}+1&1\\-m&m_{_{\rm OAM}}+1&m_i\end{pmatrix}
 \label{eq:scheme}\end{align}
(cf. appendix A for $\mathcal{E}_{-}(\varepsilon_{\rm COE})$).
These relations impose the propensity rules
$\ell - \ell_i = \Delta\ell\leq m_{\rm OAM}+1\, \mbox{ for }\,    \ell_i+\ell+m_{\rm OAM}\,  \mbox{ is odd,}$  and $\, m-m_i=\Delta m=m_{\rm OAM}+1.$
  Photoelectrons  originating from $m_i=0$ avoid  the $x-y$ plane (i.e. $\vartheta_{\boldsymbol{k}}=\pi/2$) since the spherical harmonics $Y_{\lambda,\lambda-1}(\Omega_{\boldsymbol{k}})$  have a node at $\vartheta_{\boldsymbol{k}}=\pi/2$. The emission probability  $|a_{ i}(\boldsymbol{k})|^2$ exhibits no angular dependence in the equatorial plane.
Around the Cooper minimum transitions to lower orbital angular momenta are weaker  \cite{cooper1962photoionization}.
The energetic position of the minimum depends strongly  on the angular momentum of the perturbative field. Fig.~\ref{fig1}(a)  shows the radial matrix elements for $m_{_{\rm OAM}}=1$. The relevant transitions according to the scheme~\eqref{eq:scheme} are the transitions $\ell_{ i}=1\rightarrow\ell=3$ and $\ell_{ i}=1\rightarrow\ell=1$. Around a laser frequency of $\omega=95$\,eV we find that the expectedly  dominant $d_{\ell=3,\ell_{ i}=1}^{m_{_{\rm OAM}}=1}$ has a comparable magnitude as $d_{\ell=1,\ell_{ i}=1}^{m_{_{\rm OAM}}=1}$.  {A strong angular dependence of the time delay is expected  in the energy regime where the strengths of both ionization channels are comparable, for   the interference between both channels delivers eventually the angular modulation \cite{watzel2015angular}. This motivates
 our choice of the frequency regime, both for Ar and C$_{60}$.}
The underlying physics of  time delay both for Ar and C$_{60}$ is similar, and we will elaborate  here on  Ar  deferring  C$_{60}$ case to appendix F.
For $\omega=100$\,eV the Ar photoionization probability  dependence on the photoelectron emission angle $\vartheta_{\boldsymbol{k}}$ is shown in Fig.~\ref{fig2} for  different initial states $m_{ i}$.
For a topological charge $m_{_{\rm OAM}}=1$ and  $m_{ i}=1$, the ionized electron ends up in the $f$-partial wave channel with $m=3$, while the counter-rotating photoelectron ends up in a superposition of the $p$ and $f$ partial wave channels with $m=1$. A photoelectron lauched from  $m_{ i}=0$ is described by the $f$ partial wave channel with $m=2$, i.e. the node of the spherical harmonic $Y_{3,2}(\vartheta_k=\pi/2,\varphi_k)$ leads to vanishing  emission  in this direction. In the $x-y$ plane ($\vartheta_{\boldsymbol{k}}=\pi/2$) the co-rotating electron with $m_{ i}=1$ relative to the circularly polarized OAM-field is dominant over the counter-rotating one with $m_{ i}=-1$. The electron with $m_i=0$  does not escape  in this direction. Interestingly, the two types of electrons are predominantly emitted in different directions {(at $\vartheta_{\boldsymbol{k}}=150^\circ$ the counter-rotating electron  dominates  the co-rotating one)}  allowing thus a discrimination via angular resolved photoelectron detection. \\
%{This is in contrast to conventional linearly polarized light beams where  photoelectrons originating from initial states with $m_i=\pm m$ %have the same angular dependence.}\\
%
\section{Attosecond time delay}
The Wigner time delay is
% given
%by the energy variation of the phase of the photoionization amplitude
\begin{align}
%\begin{eqnarray}
\tau_{\rm W}^{i}(\varepsilon_{k},\Omega_{\boldsymbol{k}})=\frac{\partial}{\partial\varepsilon_k}\mu_{ i}(\varepsilon_{k},\Omega_{\boldsymbol{k}}), \mbox{ where } \mu_{ i}(\varepsilon_{k},\Omega_{\boldsymbol{k}})={\rm arg}\left[a_{ i}(\boldsymbol{k})\right], \nonumber\\
%\label{eq:TD1}\\
%\end{align}
%where . Alternatively  we may write
%
\mbox{ or }\quad \tau_{\rm W}^{i}(\varepsilon_{k},\Omega_{\boldsymbol{k}})=\Im\left[\frac{1}{a_i(\boldsymbol{k})}\frac{\partial a_i(\boldsymbol{k})}{\partial\varepsilon_k}\right].
\label{eq:TD2}
\end{align}
%\end{eqnarray}
The analytical expressions for $\partial a(k,\Omega_{\boldsymbol{k}})/\partial\varepsilon_k$ are  in appendix D.
  {Evaluating eq.\,\eqref{eq:TD2} on the energy shell $\varepsilon_k=\varepsilon_{\rm COE}$ reveals angular modulations with  the azimuthal angle of the form  $\exp[i(2m_{_{\rm OAM}}+2)\varphi]$, while the amplitude of this modulation depends on $\left.\partial\mathcal{E}_{+}/\partial\varepsilon_{k}\right|_{\varepsilon_k=\varepsilon_{_{\rm COE}}}$.}
  $\left.\partial\mathcal{E}_{+}/\partial\varepsilon_{k}\right|_{\varepsilon_k=\varepsilon_{_{\rm COE}}}$ depends on the pulse length (number of optical cycles $n$) and  these variations diminish quickly for longer the pulses.
\begin{figure*}[t!]
\centering
\includegraphics[width=17.0cm]{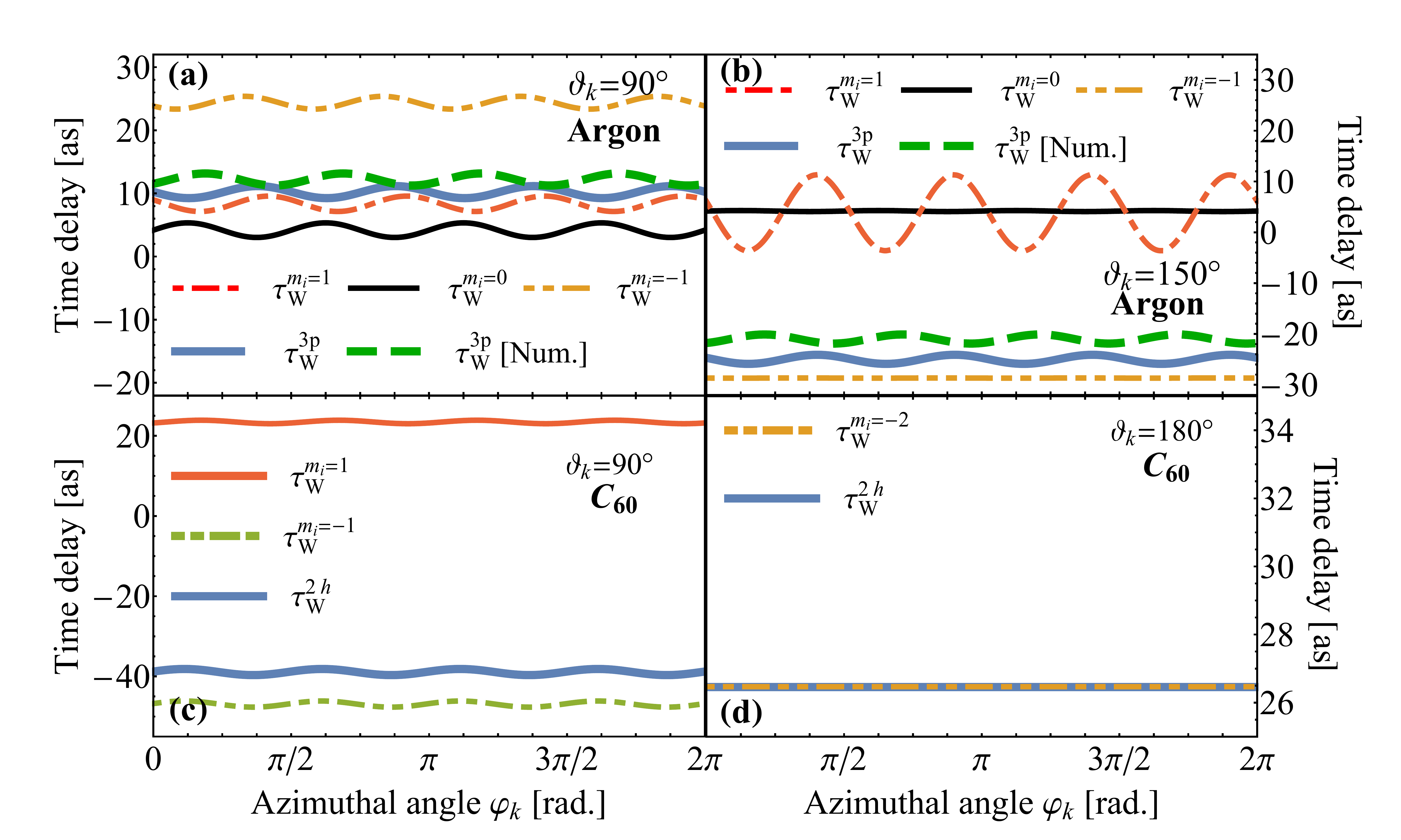}\vspace*{0.2cm}
\caption{Time delays for vortex beam  ionization
 of $3p$ subshell of Ar (a and b) as a function of  $\varphi_{\boldsymbol{k}}$ for different $\vartheta_{\boldsymbol{k}}$.
 %a)  For $\vartheta_{\boldsymbol{k}}=90^{\circ}$ (cross section is dominated by electrons with $m_{ i}=-1$), while for b)  $\vartheta_{\boldsymbol{k}}=90^{\circ}$ (electrons with $m_{ i}=1$ dominate the cross sections).
 (c,d) time delay for the vortex pulse ionization of the highest occupied molecular orbitals of C$_{60}$ fullerenes.
In (c) $\vartheta_{\boldsymbol{k}} = 90^{\circ}$ (electrons with $m_i =-1$ dominate); in (d)
 $\vartheta_{\boldsymbol{k}}= 180^{\circ}$ (mostly electrons from  $m_i = -2$ are emitted). Full averaged time delays
for the $5h$ subshell are shown.}
\label{fig2}
\end{figure*}
The time delay associated with  a  subshell averaged over $m_{ i}$ is
\begin{align}
\tau_{\rm W}^{n_{ i}\ell_{ i}}(\Omega_{\boldsymbol{k}})=\frac{\sum_{m_{ i}=-\ell_{ i}}^{\ell_{ i}} w_{\ell_im_{ i}}(\varepsilon_{\rm COE},\Omega_{\boldsymbol{k}})\tau_{\rm W}^{\ell_im_{ i}}(\varepsilon_{\rm COE},\Omega_{\boldsymbol{k}})}{\sum_{m_{ i}=-\ell_{ i}}^{\ell_{ i}}w_{\ell_im_{ i}}(\varepsilon_{\rm COE},\Omega_{\boldsymbol{k}})}.
\label{eq:TD_Full}
\end{align}
In addition to this quasi analytical model  we  solved the three-dimensional Schr\"{o}dinger equation numerically using the matrix iterative method \cite{nurhuda1999numerical,grum2010ionization}. This numerical  algorithm was alreadly  tested and implemented in time delay calculations \cite{ivanov2011time,ivanov2013time} (cf. appendix E for details).
%
%
% Here the time dependent wave function is expanded in spherical harmonics, i.e. $\Psi(\boldsymbol{r},t)=\sum_{\ell=0}^{L_{\rm max}}\sum_{m=-\ell}^{\ell}R_{\ell}(r)Y_{\ell m}(\Omega_{\boldsymbol{r}})$ with  $\lim_{t\rightarrow-\infty} \Psi(\boldsymbol{r},t)=\Psi_{ i}(\boldsymbol{r},t)$. Every initial state of  Ar 3p subshell is propagated from $t=-0.5T$ to $0.5T$ in the presence of the OAM laser field. At a time where the photoelectron wave packet is fully formed, the solution $\Psi(\boldsymbol{r},t>0.5T)$ is then projected onto a set of field-free scattering wave function $\varphi_{\boldsymbol{k}}^{(-)}(\boldsymbol{r})$ and we obtain the photoionization amplitudes $a_i(\boldsymbol{k})$ [cf. eq.\,\eqref{eq:acoeff}] associated with the specific initial state $i$, which are further analyzed to extract the time delay  [cf. eq.\,\eqref{eq:TD1}].\\
%
%
%
%
The time delays  in fig.~\ref{fig2} show, depending on the emission direction, a large difference between the photoionization process from initial states with $m_{ i}=1$ and $m_{ i}=-1$s. The photoelectron originating from $m_{ i}=1$ dominates the photoionization probabilities (cf.~fig.~\ref{fig1})  at the angle $\vartheta_{\boldsymbol{k}}=90^\circ$, while at $\vartheta_{\boldsymbol{k}}=150^\circ$ the counter-rotating electron ($m_{ i}=-1$) delivers the largest contribution.
%Dependence of the time delay on the pulse duration (or optical cycles $n$)  is also shown.
%
The small angular variations in the time delay (in $\varphi_{\boldsymbol{k}}$) smoothen very fast (without affecting the magnitude of the time-delay) for longer pulses (cf. appendix C-E).
Experimentally advantageous is the large difference between both cases where the co-rotating or the counter-rotating electrons dominate the photoionization process. The averaged time delay $\tau_{\rm W}^{3p}(\vartheta_{\boldsymbol{k}}=90^\circ)=10.7$\, \emph{as} which coincides almost with the value of $\tau_{\rm W}^{m_{ i}=1}=8.7$ \emph{as}. The time delay $\tau_{\rm W}^{m_{ i}=-1}$ being related to  the counter-rotating electron is only a minor contribution to the full subshell delay due to the lower photoionization probability.
 The electron ionized from the initial state with $m_{ i}=0$ has no influence on the resulting time delay because we find no photoionization probability in the equatorial plane. The differences between the analytical model and the numerical propagation are vanishingly small  giving further  credibility to  the analytical explanations. \\
In contrast at $\vartheta_{\boldsymbol{k}}=150^\circ$ the fully averaged, subshell time delay $\tau_{\rm W}^{3p}=-23.5$ \emph{as} is mainly characterized by $\tau_{\rm W}^{m_{ i}=-1}=-27$ \emph{as}, where the influences of the co-rotating electron ($\tau_{\rm W}^{m_{ i}=+1}=3.0$ \emph{as}) and the electron ionized from the initial state with $m_{ i}=0$ ($\tau_{\rm W}^{m_{ i}=0}=4.0$ \emph{as}) play a minor role. Thus, we find a large difference of 34.2 \emph{as} between  both cases where either the co-rotating ($m_{ i}=1$) electron or the counter-rotating electron ($m_{ i}=-1$) dominates. With this configuration it is so possible to pinpoint the origin of the time delay, i.e.~a time delay measurement  identifies from which initial magnetic sublevel the photoelectron were lauched. From the analytical and symmetry considerations it is conceivable that these findings are of a general nature  and are akin to quantized systems with spherical symmetry.
This is indeed confirmed by  corresponding results  (Fig.\ref{fig2}c,d) for  ionization of C$_{60}$ from the highest occupied molecular levels (HOMO) (see appendix F for full technical details).  The 5 electrons in HOMO (or the $5h$ state) occupy the  magnetic sublevels $m_i=\pm 2,\pm 1, 0$ which are
degenerate but their
photoionization probabilities  exhibit crossly different angular behavior, as  for Ar:
In certain directions  the photoionization  is
 dominated by emission from  specific  initial  magnetic sublevels of HOMO. As a result, if for instance
 $\vartheta_{\boldsymbol{k}}=90^\circ$ or
$\vartheta_{\boldsymbol{k}}=180^\circ$ are chosen where photoionization stems largely from $m=-1$ or $m=-2$ respectively,  we observe the azimuthal time delay behavior as depicted in
Fig. \ref{fig2}c,d.   The interpretation goes along the lines as for Ar.
 The time delays averaged over the initial degeneracies  are governed by contributions from the $m_i$ states that dominate the photoionization.\\
\section{Spatial dependence of time delay}
Another interesting aspect is the dependence of the time delay in  photoionization  on the position $r_0$  of the atom in the OAM XUV laser spot, i.e. away from the optical axis.
 When the atom is in the donut center the transfer of  OAM  from the light beam to the photoelectron is maximal, decreasing with enhancing  the distance between the atom and optical axis $r_0$ \cite{Watzel2016Driving}. This is  due to the vast difference in the spatial extension of the atom and the laser spot. Roughly speaking, when the atom is at the peak intensity ($r_0\approx w_0/\sqrt{2}$) only the beam local spatial structure   is relevant, which resembles locally  a Gaussian beam \footnote{Note3}.
\begin{figure}[t!]
\centering
\includegraphics[width=8.5cm]{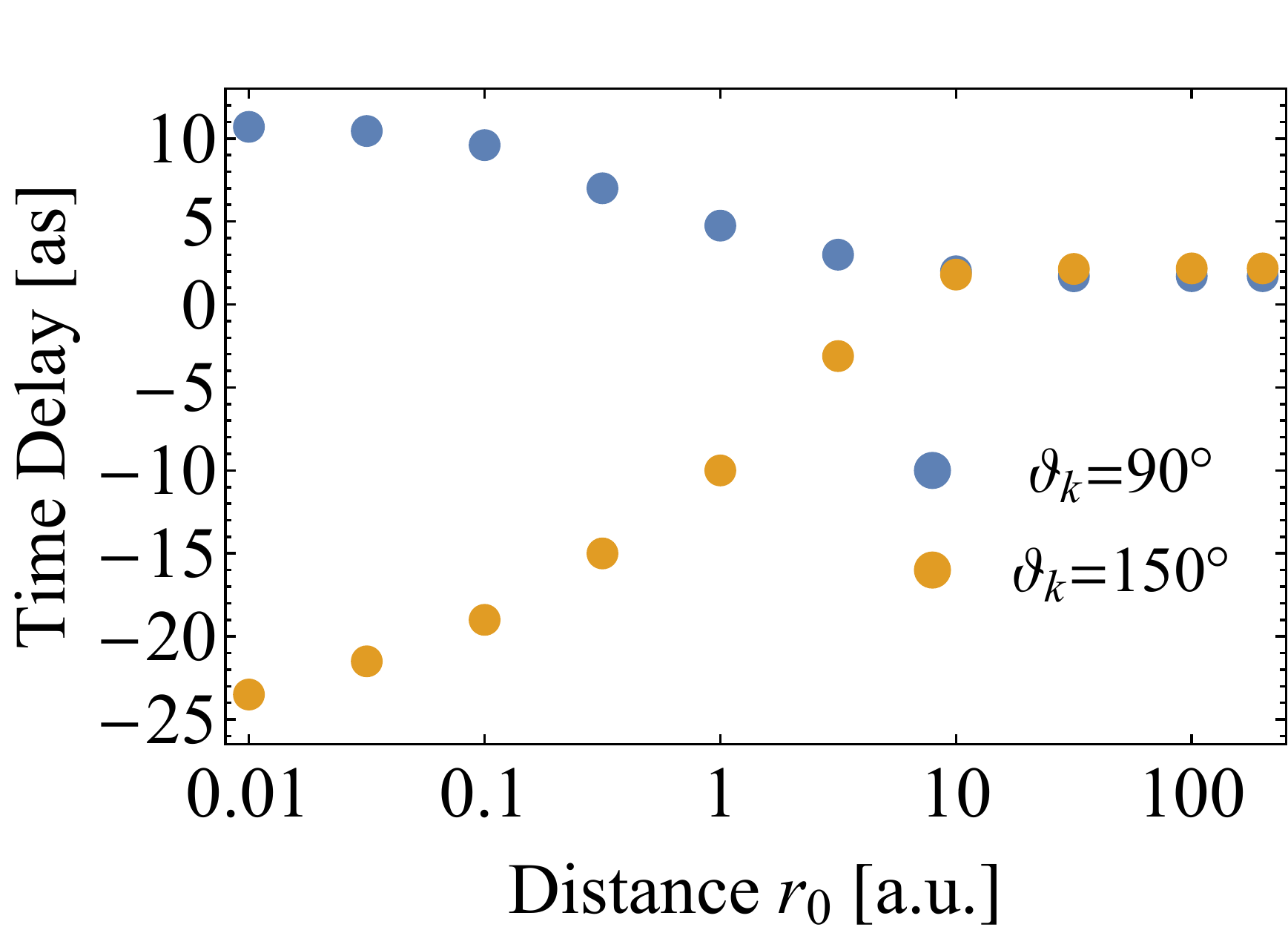}
\caption{Time delay  variation with the atom distance $r_0$ from the optical axis. Pulse duration is $n=10$ optical cycles. Other pulse parameters are as in Fig.(\ref{fig2}).}
\label{fig3}
\end{figure}
In fig.\,\ref{fig3} we show the time delay corresponding to  cases where either the co-rotating electron ($\vartheta_{\boldsymbol{k}}=90^\circ$) or the counter-rotating electron ($\vartheta_{\boldsymbol{k}}=150^\circ$)
dominates the photoionization process, as delivered from the full numerical simulations.
 The numbers at $r_0=0$ belong to the results of fig.\,\ref{fig2}. Surprisingly, even at small distances $r_0\approx1$\,a.u. the transfer of OAM diminishes rapidly. {At distances $r_0>10$\,a.u. both time delays are nearly indistinguishable}. So we argue that measurements of the time delay as a function of the topological charge  allows accessing magnetic information with an
 atomic size spatial resolution using optical beams. This is not a violation of the diffraction limit, as these information are carried by the photoelectron and are not gained via optical microscopy.
\section{Conclusions}
Summarizing, the time delays in photoionization are substantially different for  co-rotating  (relative to the OAM field) or  counter-rotating emitted electrons, even for spherically  symmetric targets.  The time delay carry  atomic-scale  information on the  orbital position in the beam spot. Including spin-orbital coupling, e.g., as done  in \cite{quinteiro2011orbital} should yield spin-dependent time delays offering  a tool for polarized electron burst \cite{clayburn2013search} by short OAM pulses. Combined with the possible spatial resolution on the magnetic states, this may offer a novel technique for spatio-temporal mapping of spin dynamics.\\
\acknowledgments The work is supported by the DFG through SPP 1840.
We thank Olga Smirnova and Ingo Barth for interesting discussions.

\appendix

\section{Modelling propagating OAM beams}
The OAM beam vector potential  in the coordinate frame of the atom with the $z$ axis being parallel to the light propagation (with a wave vector $q_z$) is  \cite{allen1992orbital}
\begin{equation}
\boldsymbol{A}(\boldsymbol{r},t)=\hat{\epsilon}A_0f_{m_{_{\rm OAM}}}^p(\boldsymbol{r}) e^{i\left[m_{_{\rm OAM}}\varphi'(\boldsymbol{r})-\omega t\right]}\, g(t)e^{iq_z z'(\boldsymbol{r})} + {\rm c.c.}.
\end{equation}
  The polarization vector is taken $\hat{\epsilon}=(1,i)^T$,
 and  for the pulse temporal envelope we take
 $$ g(t) = \cos[\pi t/nT]^2, $$ where $T=2\pi/\omega$ is the cycle duration and $n$ is the number of optical cycles. $\varphi'(\boldsymbol{r})$ is the electron azimuthal angle relative to the optical axis of the laser field. If the atom is in the beam center  we write $\varphi'(\boldsymbol{r})\equiv\varphi$. For the photon energies of concern here  $q_zz'(\boldsymbol{r})\ll1$  applies, i.e. the dipole approximation is acceptable along the $z$ axis. The radial structure is described by the function
%C_{m_{_{\rm OAM}}}^p
\begin{align}
 f^p_{m_{_{\rm OAM}}}(\boldsymbol{r}) = e^{-\frac{\rho'(\boldsymbol{r})^2}{w_0^2}}\left(\frac{\sqrt{2}\rho'(\boldsymbol{r})}{w_0}\right)^{\left|m_{_{\rm OAM}}\right|}L_p^{\left|m_{_{\rm OAM}}\right|}\left(\frac{2\rho'(\boldsymbol{r})^2}{w_0^2}\right),
 \end{align}
where $\rho'(\boldsymbol{r})$ is the radial distance to the optical axis. If the atom is in the beam center  then $\rho'(\boldsymbol{r})=r\sin\vartheta$.
 The number of nodes in the  beam radial profile is indexed by $p$ and
 $L_p^{\left|m_{_{\rm OAM}}\right|}(x)$ are the generalized Laguerre polynomials. We consider the experimentally
important case $p=0$ for which  $L_p^{\left|m_{_{\rm OAM}}\right|}(x)=1$.
   Calculations for $p\neq0$ are  feasible but are not expected to yield any sizable effect on the time delay (the beam radial variation is on the scale of tens of nanometers, i.e. far off the electron wavelength).
   $w_0$ stands for the beam waist. Typical values that we employed in the calculations  are in the range of
 $w_0=50$\,nm (940\,a.u.).
  Obviously $|\mathbf A|^2$  possess a donut shape for $p=0$ and intercalated rings for $p>1$.
\section{Transition amplitude}
For the analytical model and the situation detailed in the main text
 the optical-vertex matrix elements between the initial and final states are (we exploited $\boldsymbol{p}=-\left[\hat{H}_0,\boldsymbol{r}\right]_-$)
%\begin{equation}
$\langle \varphi_{\boldsymbol{k}}^{(-)}|\hat{H}_{int}(t)|\Psi_{ i}\rangle=i(\varepsilon_{ i}-\varepsilon_{\boldsymbol{k}})\langle \varphi_{\boldsymbol{k}}^{(-)}|\boldsymbol{r}\cdot\boldsymbol{A}(\boldsymbol{r},t) |\Psi_{ i}\rangle.
$
%\end{equation}
After the laser pulse is  off we infer
\begin{equation}
\begin{split}
a_i(\boldsymbol{k})=&(\varepsilon_{ i}-\varepsilon_{\boldsymbol{k}})\sum_{\ell=0}\sum_{m=-\ell}^{m=\ell}i^{-\ell}e^{i\delta_{\ell}(k)}d_{\ell,n_{ i}\ell_{ i}}^{m_{_{\rm OAM}}}Y_{\ell m}(\Omega_{\boldsymbol{k}})\\
&\times\left[\mathcal{E}_{-}(\varepsilon_{\boldsymbol{k}}-\varepsilon_{ i})\begin{pmatrix}\ell&m_{_{\rm OAM}}+1&\ell_i\\-m&m_{_{\rm OAM}}+1&m_i\end{pmatrix} \right.\\
&\left.+ \mathcal{E}_{+}(\varepsilon_{\boldsymbol{k}}-\varepsilon_{ i})\begin{pmatrix}\ell&m_{_{\rm OAM}}+1&\ell_i\\-m&-m_{_{\rm OAM}}-1&m_i\end{pmatrix}\right],
\end{split}
\end{equation}
where
$\mathcal{E}_{\mp}(\varepsilon)=\mathcal{E}_0\int_{-\infty}^{\infty}{\rm d}t\,g(t)e^{ i(\varepsilon\mp\omega)t}$  and $\mathcal{E}_0=A_0\left(\frac{\sqrt{2}}{w_0}\right)^{\left|m_{_{\rm OAM}}\right|}$. The {reduced} radial matrix elements are given by
\begin{equation}
\begin{split}
d_{\ell,n_{ i}\ell_{ i}}^{m_{_{\rm OAM}}}=&\sqrt{\frac{(2\ell+1)(2m_{_{\rm OAM}}+3)(2\ell_{ i}+1)}{3}}\\
&\times\begin{pmatrix}\ell&m_{_{\rm OAM}}+1&\ell_i\\0&0&0\end{pmatrix}\int{\rm d}r\,r^{3+m_{_{\rm OAM}}}R_{k\ell}(r)R_{ i}(r).
\end{split}
\label{eq:RadialMatrix}
\end{equation}
\section{Details to the numerical propagation scheme}
 Numerically, we follow a standard  matrix iterative method: The time dependent wave function is expanded in spherical harmonics, i.e. $\Psi(\boldsymbol{r},t)=\sum_{\ell=0}^{L_{\rm max}}\sum_{m=-\ell}^{\ell}R_{\ell}(r)Y_{\ell m}(\Omega_{\boldsymbol{r}})$ with  $\lim_{t\rightarrow-\infty} \Psi(\boldsymbol{r},t)=\Psi_{ i}(\boldsymbol{r},t)$. Every initial state of  Ar 3p subshell is propagated from $t=-0.5T$ to $0.5T$ in the presence of the OAM laser field. At a time where the photoelectron wave packet is fully formed, the solution $\Psi(\boldsymbol{r},t>0.5T)$ is then projected onto a set of field-free scattering wave function $\varphi_{\boldsymbol{k}}^{(-)}(\boldsymbol{r})$ and we obtain the photoionization amplitudes $a_i(\boldsymbol{k})$  associated with the specific initial state $i$, which are further analyzed to extract the time delay.\\
\section{Time Delay}
The  Wigner time delay in photoionization  is  given by
\begin{equation}
\tau_{\rm W}^{i}(\varepsilon_{k},\Omega_{\boldsymbol{k}})=\frac{\partial}{\partial\varepsilon_k}\mu_{ i}(\varepsilon_{k},\Omega_{\boldsymbol{k}}),
\end{equation}
where $\mu_{ i}(\varepsilon_{k},\Omega_{\boldsymbol{k}})={\rm arg}\left[a_{ i}(\boldsymbol{k})\right]$ or  by
\begin{equation}
\tau_{\rm W}^{i}(\varepsilon_{k},\Omega_{\boldsymbol{k}})=\Im\left[\frac{1}{a_i(\boldsymbol{k})}\frac{\partial a_i(\boldsymbol{k})}{\partial\varepsilon_k}\right].
\end{equation}
Taking into account that $\partial\mathcal{E}_{-}/\partial{\varepsilon_{k}}=0$ (absorption) while $\partial\mathcal{E}_{+}/\partial{\varepsilon_{k}}\neq0$ (emission) at $\varepsilon_k=\varepsilon_{_{\rm COE}}$, we find the following expression for the energy derivative of the amplitude in case of $m_{ i}=1$
\begin{equation}
\begin{split}
\left.\frac{\partial a(k,\Omega_{\boldsymbol{k}})}{\partial\varepsilon_k}\right|_{\varepsilon_k=\varepsilon_{_{\rm COE}}}=&\frac{\partial S_{m_{_{\rm OAM}}+2,m_{_{\rm OAM}}+2}}{\partial\varepsilon_k}Y_{m_{_{\rm OAM}}+2,m_{_{\rm OAM}}+2} (\Omega_{\boldsymbol{k}})\\
&+F_{m_{_{\rm OAM}},-m_{_{\rm OAM}}}Y_{m_{_{\rm OAM}},-m_{_{\rm OAM}}}(\Omega_{\boldsymbol{k}})\\
&+F_{m_{_{\rm OAM}}+2,-m_{_{\rm OAM}}}Y_{m_{_{\rm OAM}}+2,-m_{_{\rm OAM}}}(\Omega_{\boldsymbol{k}})
\end{split}
\end{equation}

\begin{figure*}[t!]
\includegraphics[width=17cm]{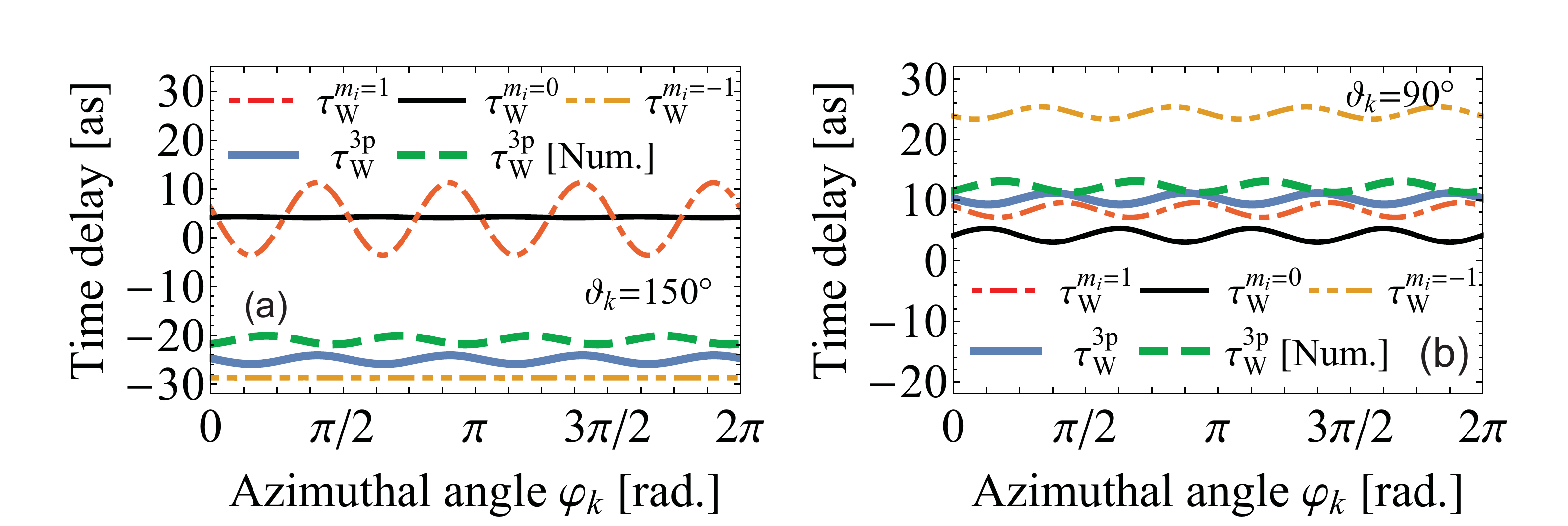}
\includegraphics[width=17cm]{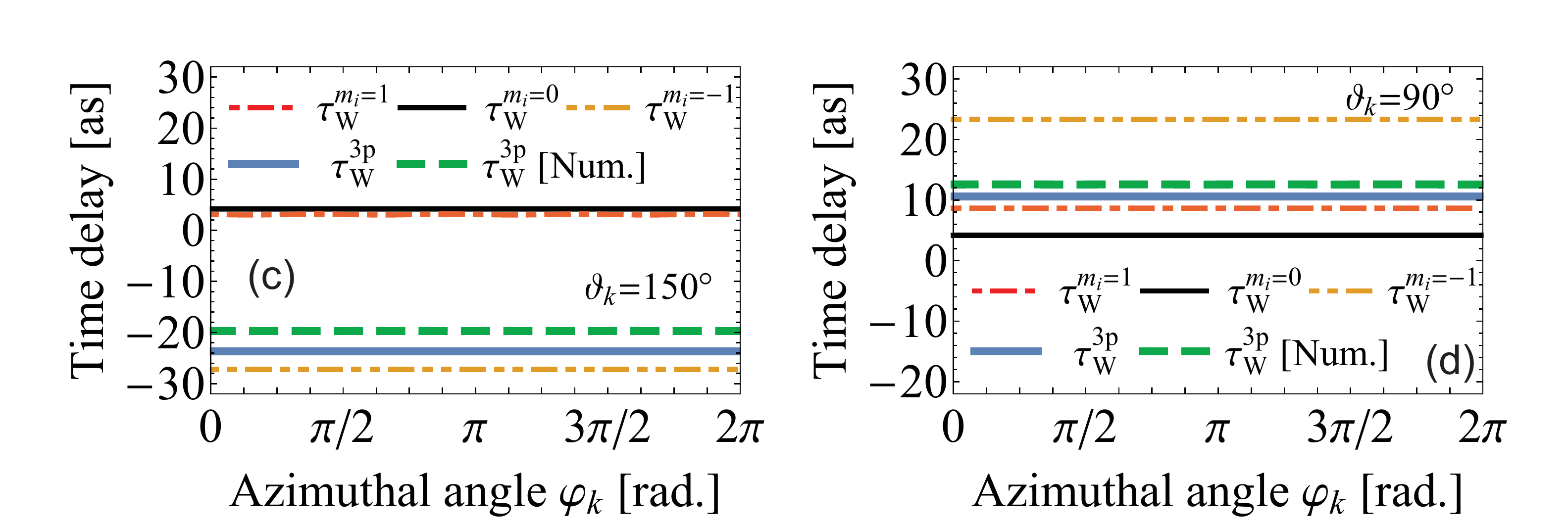}
\caption{Dependence of the time delays on the pulse duration. This figure is to be compared with Fig.2 of the main text.  For the  upper panels the pulse consists of $n=3$ optical cycles, while for the lower panels the pulse has  $n=10$. Other pulse parameters are the same as in Fig.2 of the main text. For $\vartheta_k=150^\circ$ the averaged time delay stems from states with $m_i=-1$, whereas for  $\vartheta_k=90^\circ$ states with $m_i=+1$ deliver the major contribution to the averaged time delay (note, the ionization probability for $m_i=-1$ in  this region is suppressed).}
\label{fig:fig1}
\end{figure*}

where
\begin{equation}
F_{\ell,m}=\left.\frac{\partial\mathcal{E}_{+}}{\partial\varepsilon_k}d_{\ell,n_{ i}\ell_{ i}}^{m_{_{\rm OAM}}}i^{-\ell}(k)e^{i\delta_{\ell}(k)}\begin{pmatrix}\ell&m_{_{\rm OAM}}+1&1\\-m&-m_{_{\rm OAM}}-1&m_i\end{pmatrix}\right|_{\varepsilon_k=\varepsilon_{_{\rm COE}}}
\end{equation}
incorporates the emission coefficient. Along the same lines we obtain for $m_{ i}=0$
\begin{equation}
\begin{split}
\left.\frac{\partial a(k,\Omega_{\boldsymbol{k}})}{\partial\varepsilon_k}\right|_{\varepsilon_k=\varepsilon_{_{\rm COE}}}=&\frac{\partial S_{m_{_{\rm OAM}}+2,m_{_{\rm OAM}}+1}}{\partial\varepsilon_k}Y_{m_{_{\rm OAM}}+2,m_{_{\rm OAM}}+1} (\Omega_{\boldsymbol{k}})\\
&+F_{m_{_{\rm OAM}}+2,-m_{_{\rm OAM}-1}}Y_{m_{_{\rm OAM}}+2,-m_{_{\rm OAM}}-1}(\Omega_{\boldsymbol{k}})\\
\end{split}
\end{equation}
and for $m_{ i}=-1$
\begin{equation}
\begin{split}
\left.\frac{\partial a(k,\Omega_{\boldsymbol{k}})}{\partial\varepsilon_k}\right|_{\varepsilon_k=\varepsilon_{_{\rm COE}}}=&\frac{\partial S_{m_{_{\rm OAM}}+2,m_{_{\rm OAM}}}}{\partial\varepsilon_k}Y_{m_{_{\rm OAM}}+2,m_{_{\rm OAM}}} (\Omega_{\boldsymbol{k}})\\
&+\frac{\partial S_{m_{_{\rm OAM}},m_{_{\rm OAM}}}}{\partial\varepsilon_k}Y_{m_{_{\rm OAM}},m_{_{\rm OAM}}} (\Omega_{\boldsymbol{k}})\\
&+F_{m_{_{\rm OAM}}+2,-m_{_{\rm OAM}-2}}Y_{m_{_{\rm OAM}}+2,-m_{_{\rm OAM}}-2}(\Omega_{\boldsymbol{k}}).
\end{split}
\end{equation}
\section{Analytical vs. numerical results}

To facilitate the   comparison between the analytical and the numerical results for the delay time as the pulse duration varies
we refer to  Fig.\,\ref{fig:fig1} of this supplementary materials that should be compared with   Fig.2 of the main text.
 It is obvious that for longer pulse durations (meaning more optical cycles $n$) the small variations in the dependence
 on the azimuthal angle $\varphi_{\boldsymbol{k}}$ diminish.

\section{Time delay in photoionization of  C$_{\rm 60}$ molecule}

\begin{figure}[t!]
\centering
\includegraphics[width=7cm]{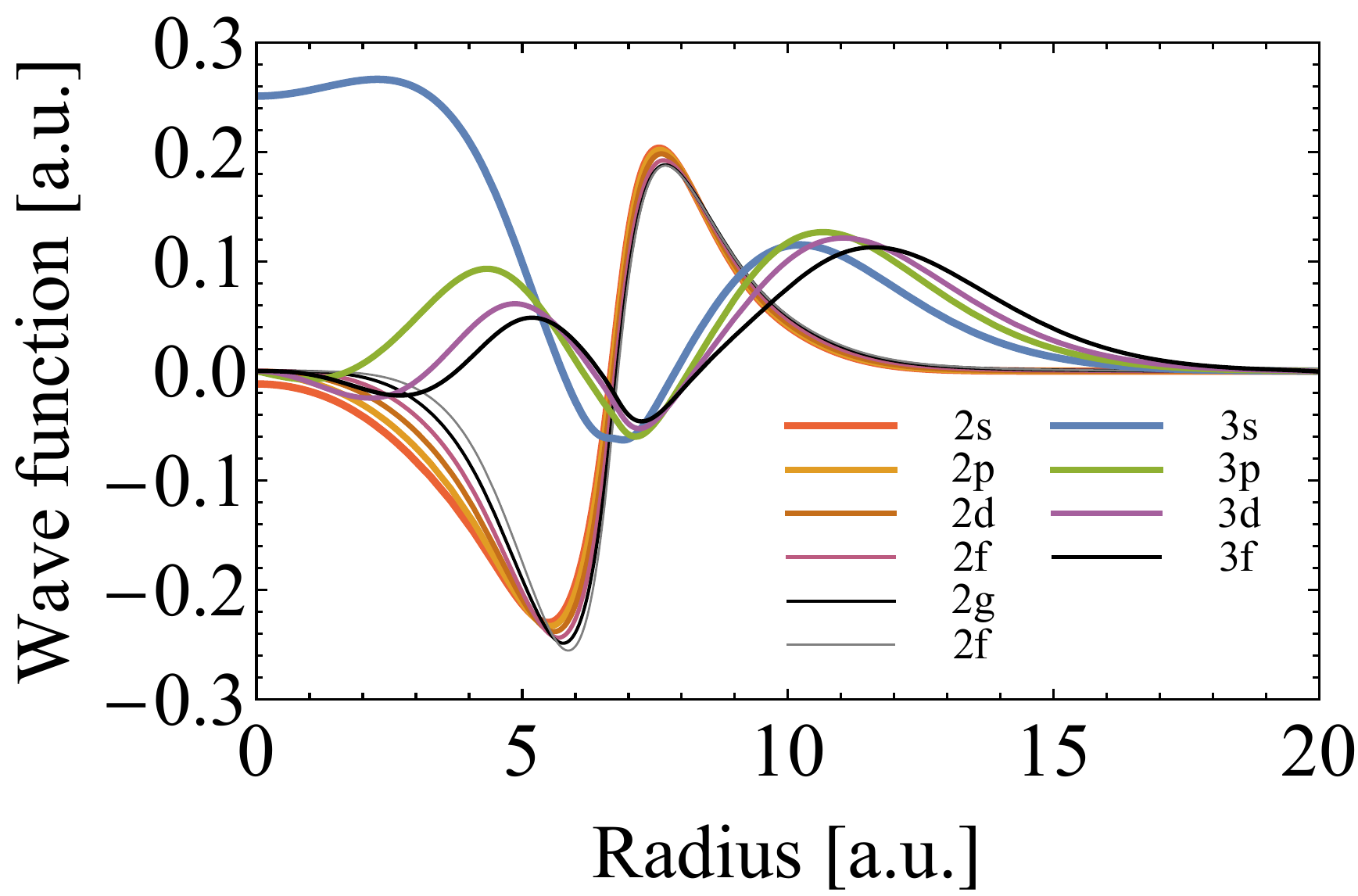}
\caption{The real radial wave functions of the electronic states for different orbital quantum numbers.}
\label{fig:wf}
\end{figure}

\begin{figure*}[t!]
\centering
\includegraphics[width=17cm]{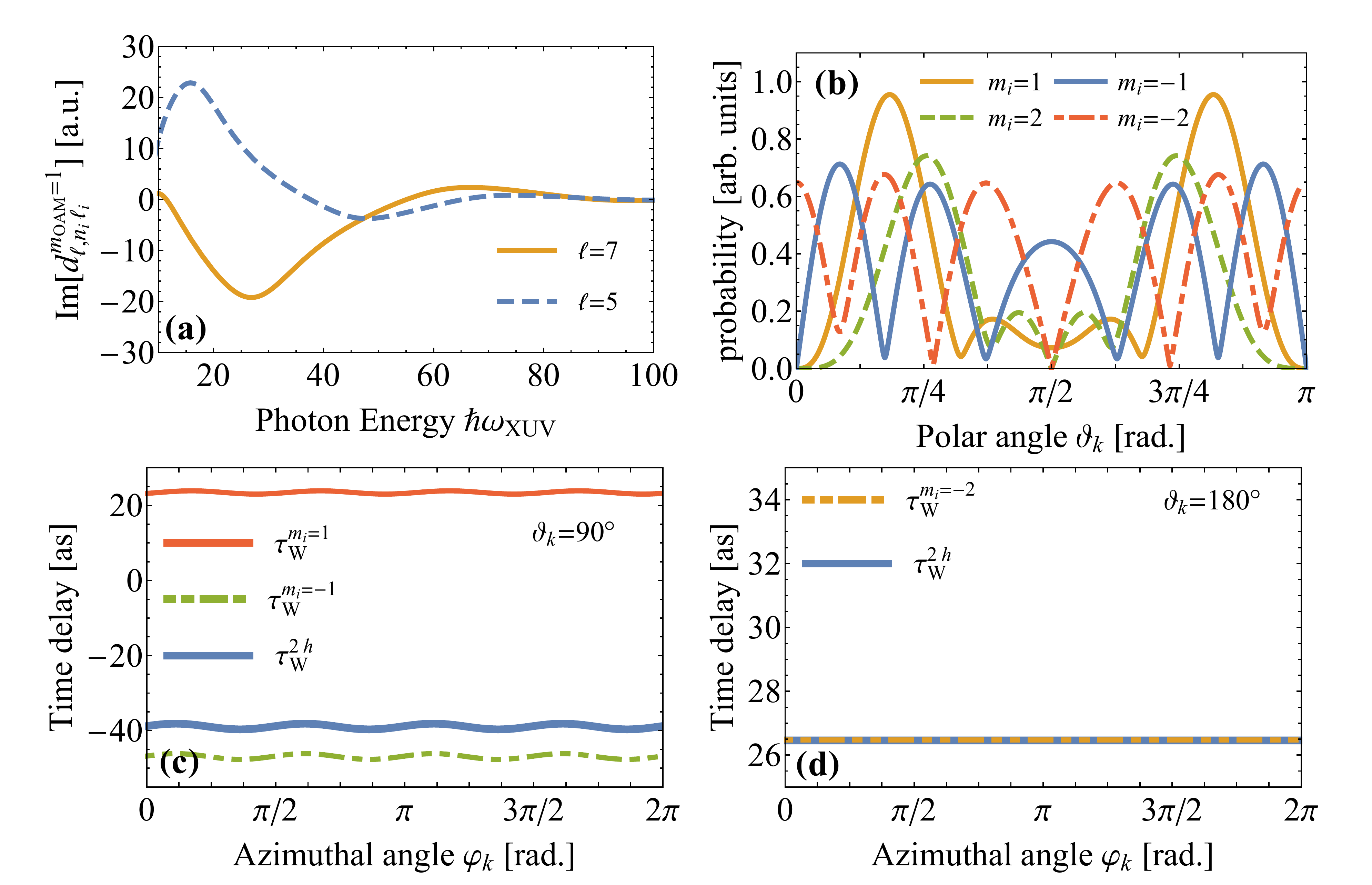}
\caption{(a) Reduced radial matrix elements for the partial wave functions with the orbital
angular momenta $\ell = 5$ and $\ell = 7$. (b) Angular dependent photoionization probabilities for the different initial states of the $5h$ subshell (HOMO) in C$_{\rm 60}$. (c)-(d) The time delays vary with the azimuthal angle $\varphi_{\boldsymbol{k}}$ at the polar angle $\vartheta_{\boldsymbol{k}}$.
The left column corresponds  to the photoionization process for $\vartheta_{\boldsymbol{k}}=90^{\circ}$ (electrons with $m_{ i}=-1$ are dominant), while the right column is associated with $\vartheta_{\boldsymbol{k}}=180^{\circ}$ (electrons with $m_{ i}=-2$ are dominant).   Time delays for  the $5h$ subshell averaged over the initial states degeneracies are also shown.}
\label{fig:PA}
\end{figure*}

Due to the vast difference between the atomic orbital extent and the focused, but diffraction limited
 laser spot the predicted effects for atoms  require highly intense laser pulses.  For instance,
the Ar calculations in the main text were performed for
 a peak intensity of $5.6\times10^{19}$~W/cm$^2$ at $w_0/\sqrt{2}$. For more extended   orbitals  similar effects in photoionization
 are achieved at lower  peak intensity which is advantageous from an experimental point of view.
  To endorse  and quantify this statement  we
  considered $C_{60}$ as the next step from atoms towards extended systems.
 The radius of the carbon cage of C$_{\rm 60}$ is $R_{\rm C_{\rm 60}}=6.745$\,a$_B$. Assuming an $A_0=0.05$\,a.u. at $R_{\rm C_{\rm 60}}$ we find a peak intensity of $3.2\times10^{17}$~W/cm$^2$ at $w_0/\sqrt{2}$. In principle, one may consider C$_{240}$ to lower the peak intensity even more.
\\
To apply our theory we describe the molecule with an effective single particle potential  that captures the valence  electronic structure with its  characteristics as derived accounting for the $I_h$-symmetry. Technically, as an input we use the  correlated, ab-initio calculated, single particle density $n(r)$ which incorporates the underlying ionic structure to construct a local single particle (orbital-dependent) potential  \cite{pavlyukh2009angular,pavlyukh2011communication,pavlyukh2010kohn}.
This potential is utilized for the driven electron dynamics \cite{pavlyukh2011communication,moskalenko2012attosecond}. Using the constructed potential, the electronic wave function $\Psi(r)$ of the fullerene valence shell is expressible as a product of a radial part $R_{n_i}(r)$ with $n_i-1$ nodes, and an angular part characterized by the spherical harmonics $Y_{\ell_i m_i}(\Omega_r)$ with the orbital and magnetic quantum numbers $\ell_i$ and $m_i$. The corresponding energies (degenerate in $m_i$) are $\varepsilon_{n_i\ell_i}$. Within this model the occupied valence states form  two radial ($\sigma$ and $\pi$) subbands. The wave functions are shown in fig.\,\ref{fig:wf}.
The occupation of the single-particle orbitals were discussed in Ref. \cite{haddon1986electronic} with the HOMO orbital ($n_i=2$,$\ell_i=5$) being occupied by 5 electrons.  C$_{\rm 60}$ has a diamagnetic character  with the HOMO magnetic sublevels  $m_i=-2,-1,0,1,2$ being populated.\\
In fig.\,\ref{fig:PA}(a) we present the radial matrix elements which are relevant for  the photoionization process of the HOMO orbital.
  For the same reason as in the main text  we choose a frequency regime where the matrix elements have similar magnitudes (in which case
   $\hbar\omega_{\rm XUV}=60$\,eV). In panel fig.\,\ref{fig:PA}(b) we show the corresponding photoionization probabilities $\left|a_{\ell_i=5,m_i}(k_{\rm COE},\vartheta_{\boldsymbol{k}})\right|^2$ of the different initial states from the $5h$ orbital in C$_{\rm 60}$ in dependence on the polar angle $\vartheta_{\boldsymbol{k}}$ relative to the optical axis of the vortex field. The figure demonstrates the significantly different angular distributions
    for photoelectrons originating from  different initial states which endorses the generality of the predicted  effect.
 Clearly, one may follow the arguments made for Ar in the main text and reach the same conclusions for C$_{\rm 60}$:
  We find directions where the photoionization process is totally dominated by some specific initial magnetic sublevel of the  HOMO.
This has also a direct consequence for the time delay depicted in panel fig.\,\ref{fig:PA}(c) and (d). We choose here as examples  the polar angles $\vartheta_{\boldsymbol{k}}=90^\circ$ and
$\vartheta_{\boldsymbol{k}}=180^\circ$, where according to the photoionization probabilities are dominated by emission from respectively the  $m_i=-1$ and $m_i=-2$ states.  The duration of the pulse is   $n=3$ optical cycles. The small variations as the azimuthal angle $\varphi_{\boldsymbol{k}}$  varies  decrease for longer pulses.
 We show only time delays of electrons which have a photoionization probability $\left|a_{\ell=5,m}(k_{\rm COE},\vartheta_{\boldsymbol{k}})\right|^2>0$. The time delay  averaged over initial state degeneracies   receives major contributions from specific magnetic sublevels
  at certain directions as in the case of  Argon atom.

\end{document}